\documentclass[12pt]{article}
\usepackage{bm}
\begin{document}
\title{NEUTRINO MASSES VIA A SEESAW WITH HEAVY MAJORANA AND DIRAC NEUTRINO MASS MATRICES FROM DISCRETE SUBGROUP $\Delta(27)$ OF $SU(3)$\footnote{Talk given on 16th International Seminar on High Energy Physics 'QUARKS 2010', Kolomna, Russia, 6-12 June 2010}}
\date{}
\maketitle
\begin{center}
\textbf{Asan Damanik}\\
\itshape Department of Physics, Faculty of Science and Technology,\\ Sanata Dharma University,\\ Kampus III USD Paingan, Maguwoharjo, Sleman,Yogyakarta, Indonesia\\
E-mail:d.asan@lycos.com\\
\end{center}

\abstract{Neutrino mass matrix via a seesaw mechanism is constructed by assumming that the underlying symmetry of both heavy Majorana and Dirac neutrino mass matrices is the discrete subgroup $\Delta(27)$ symmetry of $SU(3)$.  Using the experimental data of neutrino oscillations, the neutrino mass matrix exhibits maximal $\nu_{\mu}-\nu_{\tau}$ mixing and has a specific prediction on the effective neutrino mass in neutrinoless double beta decay which can be tested in future experiment.}

\begin{flushleft}
\end{flushleft}

\section{Introduction}
In the Standard Model of Electroweak interaction (also known as Glashow-Weinberg-Salam model), neutrinos (as a Dirac particle) are massless due to nature of neutrinos it self which only left-handed neutrinos or right-handed antineutrinos participate in the weak interactions.  Recently, there is a convincing evidence that neutrinos have a non-zero mass.  This evidence was based on the experimental facts that both solar and atmospheric neutrinos undergo oscillations \cite{Fukuda98,Fukuda99, Ahn03,Toshito01, Giacomelli01, Ahmad02}.  The neutrino oscillation implies that the neutrinos have a non-zero mass and mixing does exist in neutrino sector.

A global analysis of neutrino oscillations data gives the best fit value to solar neutrino squared-mass differences \cite{Gonzales-Garcia04},
\begin{eqnarray}
\Delta m_{21}^{2}=(8.2_{-0.3}^{+0.3})\times 10^{-5}~{\rm eV^2}~
 \label{11}
\end{eqnarray}
with
\begin{eqnarray}
\tan^{2}\theta_{21}=0.39_{-0.04}^{+0.05},
 \label{21}
\end{eqnarray}
and for the atmospheric neutrino squared-mass differences
\begin{eqnarray}
\Delta m_{32}^{2}=(2.2_{-0.4}^{+0.6})\times 10^{-3}~{\rm eV^2}~
 \label{22}
\end{eqnarray}
with
\begin{eqnarray}
\tan^{2}\theta_{32}=1.0_{-0.26}^{+0.35},
\end{eqnarray}
where $\Delta m_{ij}^2=m_{i}^2-m_{j}^2~ (i,j=1,2,3)$ with $m_{i}$ is the neutrino mass in eigenstates basis $\nu_{i}~(i=1,2,3)$, and $\theta_{ij}$ is the mixing angle between $\nu_{i}$ and $\nu_{j}$.  The mass eigenstates basis are related to the weak (flavor) eigenstates basis $(\nu_{e},\nu_{\mu},\nu_{\tau})$ as follows,
\begin{eqnarray}
\bordermatrix{& \cr
&\nu_{e}\cr
&\nu_{\mu}\cr
&\nu_{\tau}\cr}=V\bordermatrix{& \cr
&\nu_{1}\cr
&\nu_{2}\cr
&\nu_{3}\cr}
 \label{5}
\end{eqnarray}
where $V$ is the mixing matrix.

In accordance with the non-zero neutrino squared-mass differences and neutrino mixing, several models for the neutrino mass matrix together with the responsible mechanisms for generating it patterns have been proposed by many authors \cite{Mohapatra98,Akhmedov99,He03,Zee03,Fukugita03,Altarelli04,Dermisek04,Ma03}.  One of the most interesting mechanism that can generate a small neutrino mass is the seesaw mechanism, in which the right-handed neutrino $\nu_{R}$ has a large Majorana mass $M_{N}$ and the left-handed neutrino $\nu_{L}$ obtain a mass through leakage of the order of $~(m/M_{N})$ with $m$ is the Dirac mass \cite{Fukugita03}.

According to the seesaw mechanism \cite{Gell-Mann79}, the neutrino mass matrix $M_{\nu}$ is given by,
\begin{eqnarray}
M_{\nu}\approx -M_{D}M_{N}^{-1}M_{D}^T
 \label{Mnu}
\end{eqnarray}
where $M_{D}$ and $M_{N}$ are the Dirac and Majorana mass matrices respectively.  The mass matrix model of a massive Majorana neutrino $M_{N}$ that constrained by the solar and atmospheric neutrinos deficit and incorporating the seesaw mechanism and Peccei-Quinn symmetry have been reported by Fukuyama and Nishiura \cite{Fukuyama97}.  It has been a guiding principle that the presence of hierarchies or tiny quantities imply a certain protection symmetry in undelying physics.  The candidates of such symmetry in neutrino physics may include $U(1)_{L'}$ based on the conservation of $L_{e}-L_{\mu}-L_{\tau}=L'$ and a $\mu-\tau$ symmetry based on the invariance of flavor neutrino mass term underlying the interchange of $\nu_{\mu}$ and $\nu_{\tau}$.  As we have already known that maximal mixing in atmospheric neutrino, as well as vanishing of the $U_{e3}$, is the consequences of a $\mu-\tau$ symmetry \cite{Lam, Fuki, Ge}.  The $\mu-\tau$ symmetry also known as $2-3$ symmetry.  By evaluating the papers that have been reported so far, the application of symmetry into neutrino mass matrix is always in the neutrino mass matrix in flavor basis.  It is also pointed out by Ma \cite{Ma05} that it is more sense to consider the structure of $M_{N}$ for its imprint on $M_{\nu}$.

In order to consider the structure of the $M_{N}$ for its imprint on $M_{\nu}$, in this paper, the neutrino mass matrix $M_{\nu}$ is constructed via a seesaw mechanism with both heavy Majorana and Dirac neutrino mass matrices are assumed to have a discrete subgroup $\Delta(27)$ of $SU(3)$ as its underlying symmetry.  This paper is organized as follows: In Section 2, the discrete subgroup $\Delta(27)$ of $SU(3)$ is used as the undelying symmetry for both heavy Majorana and Dirac neutrino mass matrices.  In Section 3, a seesaw mechanism is used to obtain neutrino mass matrix and evaluate its phenomenological consequences.  Finally, the Section 4 is devoted to a conclusion.

\section{Majorana and Dirac Mass Matrices from $\Delta(27)$ Symmetry}

As previously stated in Section 1, the aim of the present Section is to construct the heavy Majorana and Dirac neutrino mass matrices based on the discrete subgroup $\Delta(27)$ of $SU(3)$ symmetry.  In order to realize the goal of this section, first we write down explicitly the irreducible representation of of $\Delta (27)$ symmetry.  The next step is to construct the heavy Majorana and Dirac neutrino mass matrices using $\Delta(27)$ symmetry with three Higgs doublets and three Higgs triplets.

As one know that the non-Abelian discrete subgroup $\Delta(27)$ has 27 elements divided into 11 equivalence classes.  It has 9 one-dimensional irreducible representations $\bf{1}_{i}$ ($i=1,2,..,9$) and 2 three-dimensional ones $\bf{3}$ and $\bf{\bar{3}}$. Its character table is given below, where $n$ is the number of elements, $h$ is the order of each element, and $\omega=$exp$(2\pi/3)$ with $1+\omega+\omega^{2}=0$.\\
\begin{center}  
Table 1: Character table of $\Delta(27)$.
\end{center}
\begin{center}
\begin{tabular}{|c|c|c|c|c|c|c|c|c|c|c|c|c|c|}\hline
$\textbf{Class}$& $n$ & $h$ & $\bf{1}_{1}$ & $\bf{1}_{2}$ & $\bf{1}_{3}$ & $\bf{1}_{4}$ & $\bf{1}_{5}$ & $\bf{1}_{6}$ & $\bf{1}_{7}$ & $\bf{1}_{8}$ & $\bf{1}_{9}$ & $\bf{3}$ & $\bf{\bar{3}}$\\\hline
 $C_{1}$&$1$&$1$&$1$&$1$&$1$&$1$&$1$&$1$&$1$&$1$&$1$&$3$&$3$\\\hline
 $C_{2}$&$1$&$3$&$1$&$1$&$1$&$1$&$1$&$1$&$1$&$1$&$1$&$3\omega$ &$3\omega^{2}$\\\hline
 $C_{3}$&$1$&$3$&$1$&$1$&$1$&$1$&$1$&$1$&$1$&$1$&$1$&$3\omega^{2}$&$3\omega$\\\hline
 $C_{4}$&$3$&$3$&$1$&$\omega$&$\omega^{2}$&$1$&$\omega^{2}$&$\omega$&$1$&$\omega$&$\omega^{2}$&$0$&$0$\\\hline
 $C_{5}$&$3$&$3$&$1$&$\omega^{2}$&$\omega$&$1$&$\omega$&$\omega^{2}$&$1$&$\omega^{2}$&$\omega$&$0$&$0$\\\hline
 $C_{6}$&$3$&$3$&$1$&$1$&$1$&$\omega^{2}$&$\omega^{2}$&$\omega^{2}$&$\omega$&$\omega$&$\omega$&$0$&$0$\\\hline
 $C_{7}$&$3$&$3$&$1$&$\omega$&$\omega^{2}$&$\omega^{2}$&$\omega$&$1$&$\omega$&$\omega^{2}$&$1$&$0$&$0$\\\hline
 $C_{8}$&$3$&$3$&$1$&$\omega^{2}$&$\omega$&$\omega^{2}$&$1$&$\omega$&$\omega$&$1$&$\omega^{2}$&$0$&$0$\\\hline
 $C_{9}$&$3$&$3$&$1$&$1$&$1$&$\omega$&$\omega$&$\omega$&$\omega^{2}$&$\omega^{2}$&$\omega^{2}$&$0$&$0$\\\hline
 $C_{10}$&$3$&$3$&$1$&$\omega^{2}$&$\omega$&$\omega$&$\omega^{2}$&$1$&$\omega^{2}$&$\omega$&$1$&$0$&$0$\\\hline
 $C_{11}$&$3$&$3$&$1$&$\omega$&$\omega^{2}$&$\omega$&$1$&$\omega^{2}$&$\omega^{2}$&$1$&$\omega$&$0$&$0$\\\hline
\end{tabular}
\end{center}
The group multiplication rules are \cite{Ma06}
\begin{eqnarray}
\bf{3}\times \bf{3}=\bf{\bar{3}}+\bf{\bar{3}}+\bf{\bar{3}},
\end{eqnarray}
and
\begin{eqnarray}
\bf{3}\times\bf{\bar{3}}=\sum^{9}_{i=1}\bf{1_{i}},
\end{eqnarray}
where
\begin{eqnarray}
{\bf{1}_{1}}=1\bar{1}+2\bar{2}+3\bar{3},\ {\bf{1}_{2}}=1\bar{1}+\omega 2\bar{2}+\omega^{2}3\bar{3},\ {\bf{1}_{3}}=1\bar{1}+\omega^{2} 2\bar{2}+\omega 3\bar{3},\\
{\bf{1}_{4}}=1\bar{2}+2\bar{3}+3\bar{1},\ {\bf{1}_{5}}=1\bar{2}+\omega 2\bar{3}+\omega^{2}3\bar{1},\ {\bf{1}_{6}}=1\bar{2}+\omega^{2} 2\bar{3}+\omega 3\bar{1},\\{\bf{1}_{7}}=2\bar{1}+3\bar{2}+1\bar{3},\ {\bf{1}_{8}}=2\bar{1}+\omega^{2} 3\bar{2}+\omega 1\bar{3},\ {\bf{1}_{9}}=2\bar{1}+\omega 3\bar{2}+\omega^{2} 1\bar{3}.
\end{eqnarray}

Let the lepton doublets $(\nu_{i},l_{i})$ transform as $\bf3$ under $\Delta(27)$ and the lepton singlet $l^{c}_{i}$ as $\bar{\bf3}$, then with three Higgs doublets transforming as $\bf1_{1},\bf1_{2},\bf1_{3}$, the charged lepton and the Dirac neutrino mass matrix are diagonal and has three independent masses.  Thus, the pattern of the Dirac neutrino mass matrix as following
\begin{eqnarray}
M_{D}=\bordermatrix{& & &\cr
&a &0 &0\cr
&0 &b &0\cr
&0 &0 &c\cr}.
 \label{MD13}
\end{eqnarray}
At the same time, with three Higgs triplets transforming as $\bf3$, when vacuum expectation values of three Higgs triplets are $(\bf1,\bf1,\bf1)$, the general pattern of the heavy Majorana neutrino mass matrix is given by
\begin{eqnarray}
M_{N}=\bordermatrix{& & &\cr
&A &B &B\cr
&B &A &B\cr
&B &B &A\cr}.
 \label{M13}
\end{eqnarray}

\section{Neutrino Mass Matrix via a Seesaw Mechanism}

Using the seesaw mechanism in Eq. (\ref{Mnu}), the Dirac neutrino mass matrices in Eq. (\ref{MD13}), and the heavy Majorana neutrino mass matrices in Eq. (\ref{M13}), we finally obtain the neutrino mass matrix with pattern
\begin{eqnarray}
M_{\nu}=\bordermatrix{& & &\cr
&P &Q &R\cr
&Q &S &T\cr
&R &T &U\cr}.
 \label{MD01}
\end{eqnarray} 

If we write $P=fx$, $Q=y$, $R=z$, $S=fz$, $T=x$ and $U=fy$, then the neutrino mass matrix in Eq. (\ref{MD01}) has the form
\begin{eqnarray}
M_{\nu}=\bordermatrix{& & &\cr
&fx &y &z\cr
&y &fz &x\cr
&z &x &fy\cr}.
 \label{MD02}
\end{eqnarray} 
Given the form of Eq. (\ref{MD02}), in the limit $\theta_{13}=0$ requires $y=z$.  Within this latter assumption, the Eq. (\ref{MD02}) becomes
\begin{eqnarray}
M_{\nu}=\bordermatrix{& & &\cr
&fx &y &y\cr
&y &fy &x\cr
&y &x &fy\cr},
 \label{MD03}
\end{eqnarray}
which exhibits maximal $\nu_{\mu}-\nu_{\tau}$ mixing.  
 
One can see that the eigenvalues of the neutrino mass matrix of Eq. (\ref{MD03}) are given by
\begin{eqnarray}
\lambda_{1}=-x+fy,\label{l1}\\
\lambda_{2}=\frac{1}{2}\left[f(x+y)+x-([f(y-x)]^{2}+2fx(y-x)+x^{2}+8y^{2})^{1/2}\right],\label{l2}\\
\lambda_{3}=\frac{1}{2}\left[f(x+y)+x+([f(y-x)]^{2}+2fx(y-x)+x^{2}+8y^{2})^{1/2}\right].
 \label{l3}
\end{eqnarray}
Using experimental facts that $\Delta m_{21}^{2}<<\Delta m_{31}^{2}$, it is easy to see that the eigenvalue $\lambda_{3}$ corresponds to $m_{2}$, $\lambda_{2}$ corresponds to $m_{1}$, and $\lambda_{1}$ corresponds to $m_{3}$.  We consider first the case limit $\Delta m_{solar}^{2}=\Delta m_{21}^{2}\rightarrow 0$.  In this case, we have
\begin{eqnarray}
x\approx-\frac{fy}{f+1}.
 \label{a}
\end{eqnarray}
If we insert the value of $x$ as shown in Eq. (\ref{a}) into Eq. (\ref{MD03}), then we have neutrino mass matrix with pattern
\begin{eqnarray}
M_{\nu}\approx\bordermatrix{& & &\cr
&\frac{-f^{2}y}{f+1} &y &y\cr
&y &fy &\frac{-fy}{f+1}\cr
&y &\frac{-fy}{f+1} &fy\cr}.
 \label{MD04}
\end{eqnarray}

If neutrino mass matrix $M_{\nu}$ of Eq. (\ref{MD04}) is diagonalized by mixing matrix $V$ of Eq. (\ref{5}) with $V$ given by \cite{Ma05}
\begin{eqnarray}
V=\bordermatrix{& & &\cr
&\cos\theta &-\sin\theta &0\cr
&\sin\theta/\sqrt{2} &\cos\theta/\sqrt{2} &-1/\sqrt{2}\cr
&\sin\theta/\sqrt{2} &\cos\theta/\sqrt{2} &1/\sqrt{2}\cr},
 \label{511}
\end{eqnarray}
then we obtain,
\begin{eqnarray}
\tan\theta\approx\frac{\sqrt{2}f^{2}-\sqrt{2f^{4}+4f^{2}+8f+4}}{2(f+1)}.
 \label{Teta}
\end{eqnarray}

If $\theta$ is the $\theta_{21}$ in Eq. (\ref{21}), then from Eq. (\ref{Teta}) we can have the value of $f\approx-0.5546$ or $f\approx1.2453$.  From Eqs. (\ref{a}) and (\ref{MD04}), if we put the value of $f$ into this matrix, then the neutrino mass matrix in Eq. (\ref{MD04}) becomes
\begin{eqnarray}
M_{\nu}\approx y\bordermatrix{& & &\cr
&-0.6907 &1 &1\cr
&1 &-0.5546 &1.2453\cr
&1 &1.2453 &-0.5546\cr},
 \label{MD05}
\end{eqnarray}
for $f\approx -0.5546$, and
\begin{eqnarray}
M_{\nu}\approx y\bordermatrix{& & &\cr
&-0.6907 &1 &1\cr
&1 &1.2453 &-0.5546\cr
&1 &-0.5546 &1.2453\cr},
 \label{MD06}
\end{eqnarray}
for $f\approx 1.2453$.
It is also clear that both neutrino mass matrices in Eqs. (\ref{MD05}) and (\ref{MD06}) have the same eigenvalues which correspond to neutrino masses: $\left|m_{1}\right|\approx\left|m_{2}\right|<\left|m_{3}\right|$.  One can see that $\Delta m_{21}^{2}\approx 0$ as previously assummed. 

Using the advantages of the experimental data on neutrino oscillation as shown in Eq. (\ref{22}), from Eqs. (\ref{l1})-(\ref{l3}), for both values of $f$ we obtain the value if $y \approx 0.053705$ which then gives the neutrino mass matrix 
\begin{eqnarray}
M_{\nu}\approx\bordermatrix{& & &\cr
&-0.0371 &0.0537 &0.0537\cr
&0.0537 &-0.0298 &0.0669\cr
&0.0537 &0.0669 &-0.0298\cr},
 \label{MD07}
\end{eqnarray}
for $f\approx -0.5546$, and
\begin{eqnarray}
M_{\nu}\approx\bordermatrix{& & &\cr
&-0.0371 &0.0537 &0.0537\cr
&0.0537 &0.0669 &-0.0298\cr
&0.0537 &-0.0298 &0.0669\cr},
 \label{MD08}
\end{eqnarray}
for $f\approx 1.2453$.  From Eqs. (\ref{MD07}) and (\ref{MD08}), we obtained neutrino masses $\left|m_{1}\right|=\left|m_{2}\right|=0.0845$ eV, and $\left|m_{3}\right|=0.0967$ eV which are incompatible with the experimental data $\Delta m_{21}^{2}\neq 0$.

Now, if we consider the values of $\Delta m_{21}^{2}=8.2\times 10^{-5}$ eV$^{2}$ and $\Delta m_{31}^{2}=2.2\times 10^{-3}$ eV$^{2}$ as constraints, then we should put $\left|m_{2}\right|=0.0845$ eV in order to maintain the value of $\Delta m_{32}^{2}$, and it then implies that $\left|m_{1}\right|=0.0840$ eV.  It is apparent from above both neutrino mass matrices of Eqs, (\ref{MD07}) and (\ref{MD08}) that the effective neutrino mass $m_{ee}$ measured in neutrinoless double beta decay is simply given by the magnitude of the $\nu_{e}\nu_{e}$ entry of $M_{\nu}$, i.e. $\left|-f^{2}y/(f+1)\right|\approx 0.0371$ eV in both cases which can be tested in future experiment.

\section{Conclusion}
Neutrino mass matrix $M_{\nu}$ via a seesaw mechanism, with a discrete subgroup $\Delta(27)$ of $SU(3)$ to be assummed as the underlying symmetry for heavy Majorana and Dirac neutrino mass matrices, can explain the maximal $\nu_{\mu}-\nu_{\tau}$ mixing.  Using the experimental data as constraints for determining neutrino masses, we obtain neutrino masses in normal hierarchy: $\left|m_{1}\right|=0.0840$ eV, $\left|m_{2}\right|=0.0845$ eV, and $\left|m_{3}\right|=0.0967$ eV.  The obtained neutrino mass matrix also has a specific prediction on the effective neutrino mass in neutrinoless double beta decay which can be tested in future experiment.

\section*{Acknowledgments}
Author would like to thank to Prof. Ernest Ma for his suggestion to use the discrete subgroup $\Delta(27)$ symmetry of $SU(3)$, to Quarks 2010 Organizing Committee for a nice hospitality during the seminar, and to DP2M Ditjen Dikti Depdiknas for a financial support.

\end{document}